**Abstract**

This paper contributes to a burgeoning area of investigation, the ambiguity inherent in mathematics and the implications for physics of this ambiguity. To display the mathematical form of equations of quantum theory used to describe experiments, we make explicit the knobs by which the devices of an experiment are arranged and adjusted. A quantum description comes in two parts: (1) a statement of results of an experiment, expressed by probabilities of detections as functions of knob settings, and (2) an explanation of how we think these results come about, expressed by linear operators, also as functions of knob settings. Because quantum mechanics separates the two parts of any description, it is known that between the statements of results and the explanations lurks a logical gap: given any statement of results one has a choice of explanations.

Here we work out some consequences of this openness to choice. We show how quantum theory as mathematical language in which to describe experiments necessarily involves multiple descriptions: multiple explanations of a given result, as well as multiple statements of results and multiple arrangements of knobs. Appreciating these multiplicities resolves what otherwise is a confusion in the concept of invariance. Implications of multiplicity of description for the security of quantum key distribution are noted.

# Ambiguity in quantum-theoretical descriptions of experiments

John M. Myers[1] and F. Hadi Madjid[2]
[1]*School of Engineering and Applied Sciences, Harvard University,
60 Oxford Street, Cambridge, MA 02138*
[2]*82 Powers Road, Concord, MA 01742*

## 1 Introduction

Quantum theory can be used as a language in which to speak mathematically of particles and fields or, and this is our focus, as a language in which to describe experiments with devices, such as lasers and lenses and detectors on a laboratory bench. To employ quantum theory as mathematical language to describe experiments with devices, one assumes that the devices generate, transform, and measure particles and/or fields, expressed one way or another as linear operators. We omit discussing how one arrives at the particles, or even whether one takes them as observable or as imaginative constructs; instead we attend directly to the linear operators that end up expressing the devices. These operators are functions of the parameters by which one describes control over the devices. By making explicit the parameters that express the knobs and levers that control an experiment, we will show how quantum theory as mathematical language in which to describe experiments forces multiple descriptions. We will also show a few of the consequences of this multiplicity.

It is important to recognize that quantum theoretic descriptions of experiments come in two parts: (1) statements of results of an experiment, expressed by probabilities of detections as functions of knob settings, and (2) explanations of how one thinks these results come about, also as functions of knob settings. Given an explanation in terms of linear operators, one knows from the trace rule how to calculate the probabilities that constitute a statement of results. We will attend as much, if not more, to the "inverse problem" of choosing linear operators to explain given probabilities. When we want to create an explanation in operators of a given statement of results, these results together with the rules of linear operators act as an axiom system. From the standpoint of logic, an explanation is an interpretation of the results. Going back at least to Hilbert's bizarre interpretations of the axioms of geometry, there is a developing awareness that mathematics is ambiguous, and that this ambiguity is no fault to be repaired, but is intrinsic to mathematics and indeed to "language itself" [1].

A few years ago we proved that between the two parts of a description—the statement of results and the explanation—lurks a gap not bridged by logic, open to choice resolvable only by stepping outside logic to make an assumption that, inspired or not, can be called a guess [2]–[5]. The proof prompts further exploration, and here we report on: (1) implications of statements of results for the topology of knobs, *independent of choices of explanation*; (2) an endless cycle of extensions of both explanations and statements of results forced by openness to choice; and (3) an apparent paradox in the concept of invariance, resolved by recognizing multiple descriptions. Along the way we note implications of ambiguities of description for quantum cryptography. A take-home lesson is that descriptions expressed in quantum theory make sense only in a context of more than one description, so that relations among different descriptions—both the statements of results and their explanations—become essential ingredients in the very concept of a description.

Note that while our discussion gives knobs a prominent expression absent in text books on quantum mechanics, we employ the standard quantum mechanics of Dirac and von Neumann [6, 7], augmented only by positive-operator-valued measures, now in widespread use.

## 2 Lattices of domains of knobs and detectors

To display the dependence of quantum explanations on choices, we need first to say how to express *knobs* in the mathematical language in which we describe experimental trials, actual or anticipated. As already noted, the mathematics used to describe trials of an experiment partitions into a statement of results and an explanation, both of which depend on the knobs and levers by which one controls devices arranged into an experiment. Subsuming levers into knobs, we express any one knob by a set of *settings* of the knob. Later we will see topologies and metrics for some knobs but for the moment we take knobs just as sets. A knob depends on a level of description; what in a coarse description is "a knob" splits in some finer description into several knobs. (This ambiguity reflects the ambiguity of what to call an "element of a set"; *i.e.* an element of one set can itself be a set.)

To permit describing a given experiment at differing levels of detail and to describe several related experiments that overlap in their knobs, we introduce a lattice structure for sets of knobs (and later also a lattice for sets of detectors). We define a *knob domain* in terms of an *unordered* product of knobs, as discussed in Appendix A. Knob domains are partially ordered by the knobs they include. Given two knob domains $\boldsymbol{K}$ and $\boldsymbol{K'}$, we can form their *meet* $\boldsymbol{K} \wedge \boldsymbol{K'}$ (the knobs they share in common) and their *join* $\boldsymbol{K} \vee \boldsymbol{K'}$ (combining all the knobs involved in either), so that knob domains form a distributive lattice. When a knob domain $\boldsymbol{K}$ is an unordered product of several knobs, then an element $k \in \boldsymbol{K}$ specifies a particular setting for each of the knobs of $\boldsymbol{K}$.

A detector is expressed mathematically by a set $\Omega$ of possible outcomes. We allow for continuous detector responses by dealing with $\tilde{\Omega}$, a $\sigma$-algebra of measurable subsets of a detector $\Omega$. Just as experiments can have multiple knobs, leading to the notion of a knob domain, they can have multiple detectors, again expressed by unordered products. We call unordered products of detectors *detector domains*. Detector domains $\tilde{\boldsymbol{\Omega}}, \tilde{\boldsymbol{\Omega}}', \ldots$ form a lattice, as described in Appendix A.

# 3 Statements of results and explanations

Given a lattice of knob domains and a lattice of detector domains, let PPM denote the function that assigns to each knob domain $\boldsymbol{K}$ and each detector domain $\tilde{\boldsymbol{\Omega}}$ the set of parametrized probability measures over that pair of domains:

$$\text{PPM}(\boldsymbol{K}, \tilde{\boldsymbol{\Omega}}) \stackrel{\text{def}}{=} \{\mu | \mu \colon \boldsymbol{K} \times \tilde{\boldsymbol{\Omega}} \to [0,1]\}, \tag{1}$$

subject to the condition:

$$(\forall k \in \boldsymbol{K}) \quad \mu(k,-) \colon \tilde{\boldsymbol{\Omega}} \to [0,1] \text{ is a probability measure on } \tilde{\boldsymbol{\Omega}}. \tag{2}$$

For any given knob and outcome domains, a statement of results is some $\mu \in \text{PPM}(\boldsymbol{K}, \tilde{\boldsymbol{\Omega}})$.

## 3.1 Metric deviation of two parametrized probability measures

Here are two ways to compare parametrized probability measures. Let $\text{PrMeas}(\tilde{\boldsymbol{\Omega}})$ denote the set of probability measures on $\tilde{\boldsymbol{\Omega}}$. For any detector domain $\tilde{\boldsymbol{\Omega}}$, the Euclidean bounded metric on $[0,1]$ lifts to the uniform metric $D_{\boldsymbol{\Omega}}$ on $\text{PrMeas}(\tilde{\boldsymbol{\Omega}})$; that is for any $\nu, \nu' \in \text{PrMeas}(\tilde{\boldsymbol{\Omega}})$ we have

$$D_{\boldsymbol{\Omega}}(\nu, \nu') \stackrel{\text{def}}{=} \sup_{\omega \in \tilde{\boldsymbol{\Omega}}} |\nu(\omega) - \nu'(\omega)| = \sup_{\omega \in \tilde{\boldsymbol{\Omega}}} [\nu(\omega) - \nu'(\omega)], \tag{3}$$

where the absolute value can be dropped because a measure space is closed under complements. Applied to compare a single parametrized probability measure evaluated at two values $k_1, k_2 \in \boldsymbol{K}$, we have

$$D_{\boldsymbol{\Omega}}[\mu(k_1, -), \mu(k_2, -)] = \sup_{\omega \in \tilde{\boldsymbol{\Omega}}} [\mu(k_1, \omega) - \mu(k_2, \omega)]. \tag{4}$$

A second lift puts the uniform metric on parametrized probability measures: for $\mu_1, \mu_2 \in \text{PPM}(\boldsymbol{K}, \tilde{\boldsymbol{\Omega}})$

$$D_{\boldsymbol{K}, \boldsymbol{\Omega}}(\mu_1, \mu_2) \stackrel{\text{def}}{=} \sup_{k \in \boldsymbol{K}} \sup_{\omega \in \tilde{\boldsymbol{\Omega}}} [\mu_1(k, \omega) - \mu_2(k, \omega)]; \tag{5}$$

however a coarser way of comparing functions from sets to topological spaces can be applied across different topological spaces, and for our purpose this is the more useful comparison. With apologies to whomever knows it by another name, we call it "metric deviation" and define it in Appendix B. For $\mu$ and $\mu'$ having the same knob domain $\boldsymbol{K}$ but possibly distinct detector domains $\tilde{\boldsymbol{\Omega}}$ and $\tilde{\boldsymbol{\Omega}}'$, respectively, we define

$$\text{MetDev}(\mu, \mu') \stackrel{\text{def}}{=} \sup_{k_1, k_2 \in \boldsymbol{K}} |D_{\boldsymbol{\Omega}}[\mu(k_1, -), \mu(k_2, -)] - D_{\boldsymbol{\Omega}'}[\mu'(k_1, -), \mu'(k_2, -)]|. \tag{6}$$

An application of this metric deviation is described in Sec. 4.

## 3.2 Explanations

Besides stating results there is *explaining* them. A quantum explanation of a statement of result $\mu\colon \boldsymbol{K} \times \tilde{\boldsymbol{\Omega}} \to [0,1]$ consists of linear operators on some Hilbert space $\mathcal{H}$ as functions of the knob settings, including detection operators involving $\tilde{\boldsymbol{\Omega}}$. Products, tensor products, sums, exponentiations, etc. of operators are combined to form a triple $(\mathcal{H}, \rho, M)$ in which $\rho$ and $M$ are functions on $\boldsymbol{K}$. The function $\rho\colon \boldsymbol{K} \to \{\text{density operators on } \mathcal{H}\}$ can be called a parametrized density operator, and the function $M\colon \boldsymbol{K} \times \tilde{\boldsymbol{\Omega}} \to \{\text{Detection operators on } \mathcal{H}\}$ is a parametrized positive operator-valued measure (POVM); more precisely, for each $k \in \boldsymbol{K}$, $M(k,-)\colon \tilde{\boldsymbol{\Omega}} \to \{\text{Detection operators on } \mathcal{H}\}$ is a POVM on the measurable sets of $\boldsymbol{\Omega}$. So defined, any explanation implies a statement of results *via* the familiar trace rule

$$(\forall k \in \boldsymbol{K}, \omega \in \tilde{\boldsymbol{\Omega}}) \quad \mu(k, \omega) = \mathrm{Tr}_{\mathcal{H}}[\rho(k) M(k, \omega)], \tag{7}$$

where $\omega \in \tilde{\boldsymbol{\Omega}}$ is an outcome. Often we abbreviate this by

$$\mu = \mathrm{Tr}_{\mathcal{H}}[\rho M]. \tag{8}$$

Let Expl denote the function that assigns to each knob domain $\boldsymbol{K}$ and each detector domain $\tilde{\boldsymbol{\Omega}}$ the set of explanations over those domains:

$$\mathrm{Expl}(\boldsymbol{K}, \tilde{\boldsymbol{\Omega}}) \stackrel{\text{def}}{=} \{(\mathcal{H}, \rho, M)\}, \tag{9}$$

subject to the conditions:

1. $\rho\colon \boldsymbol{K} \to \{\text{density operators on } \mathcal{H}\}$ and

2. $(\forall k \in \boldsymbol{K})\ M(k,-)\colon \tilde{\boldsymbol{\Omega}} \to \{\text{Detection operators on } \mathcal{H}\}$ is a POVM on $\tilde{\boldsymbol{\Omega}}$.

For any such explanation, $\mathrm{Tr}_{\mathcal{H}}[\rho M] \in \mathrm{PPM}(\boldsymbol{K}, \tilde{\boldsymbol{\Omega}})$.

## 3.3 Choice of explanation

Now comes the inverse problem, with its non-uniqueness. Given a statement of results in the form of a given parametrized probability measure $\mu$, what freedom of choice is there for an explanation $(\mathcal{H}, \rho, M)$ that generates this $\mu$? Part of the answer comes as a proof of a logical ambiguity—not a break or a conflict, but a place for choice: no matter what probabilities are given as functions of knob settings, there is always room for choice of $\rho$ and $M$ [2, 4, 5].

Although we barely touch on them in this paper, mappings between domains $\boldsymbol{K} \times \tilde{\boldsymbol{\Omega}}$ and domains $\boldsymbol{K}' \times \tilde{\boldsymbol{\Omega}}'$ induce mappings from $\mathrm{PPM}(\boldsymbol{K}', \tilde{\boldsymbol{\Omega}}')$ to $\mathrm{PPM}(\boldsymbol{K}, \tilde{\boldsymbol{\Omega}})$; likewise mappings on domains induce mappings between explanations on the respective domains. If we view explanations and statements of results as two categories, the trace respects the mappings we have in mind, and so acts as a functor from explanations to statements of results. This functor is full but unfaithful.

Choices of explanations for a given statement of results arise because the trace as a functor from explanations to statements of results has a "roomy inverse," as illustrated in Fig. 1. Given $\mu \in \mathrm{PPM}(\boldsymbol{K}, \tilde{\boldsymbol{\Omega}})$, the inverse image $\mathrm{Tr}^{-1}(\mu) \subset \mathrm{Expl}(\boldsymbol{K}, \tilde{\boldsymbol{\Omega}})$ is a big set involving an infinite tower of Hilbert spaces and generically including explanations that, as we shall see, have diverse implications.

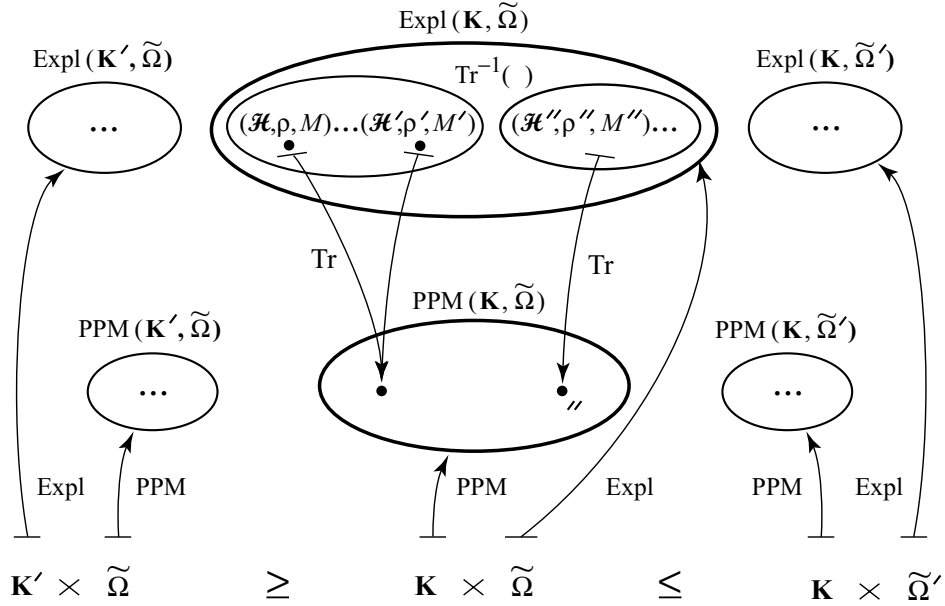

Figure 1: $\text{Tr}^{-1}(\mu)$ contains many explanations.

### 3.4 Metric deviations of explanations

The concept of metric deviation, introduced above, can be applied not only to parametrized probability measures but also to parametrized density operators and parametrized POVMs. These metric deviations of operator-valued functions of knobs allow comparisons of explanations finer-grained than a comparison of their traces; unlike operator metrics for operators on a given Hilbert space, the metric deviations allow comparisons of operators on distinct Hilbert spaces. To define the metric deviations, we first recall metrics for operators that share a common Hilbert space. Because of its role in quantum decision theory, the trace distance is the most suitable metric for positive trace-class operators, including density operators. For $\rho : \boldsymbol{K} \to \text{DensOp}(\mathcal{H})$ the trace distance between $\rho(k_1)$ and $\rho(k_2)$ is $\frac{1}{2}\text{Tr}_\mathcal{H}|\rho(k_1) - \rho(k_2)|$ [8]. For $\rho$ and $\rho'$ defined on the same domain $\boldsymbol{K}$ but with codomains $\text{DensOp}(\mathcal{H})$ and $\text{DensOp}(\mathcal{H}')$, respectively, where the Hilbert space $\mathcal{H}$ need not be the same or even isomorphic to $\mathcal{H}'$, we define a metric deviation by

$$\text{MetDev}(\rho, \rho') \stackrel{\text{def}}{=} \sup_{k_1, k_2 \in \boldsymbol{K}} \left| \tfrac{1}{2}\text{Tr}_\mathcal{H}|\rho(k_1) - \rho(k_2)| - \tfrac{1}{2}\text{Tr}_{\mathcal{H}'}|\rho'(k_1) - \rho'(k_2)| \right|. \quad (10)$$

For POVMs, the trace need not exist, and we invoke the metric derived from the norm $\|\cdot\|_\mathcal{H}$ for operators on a Hilbert space $\mathcal{H}$ [9]. For two POVMs with detection operators for outcomes in $\tilde{\boldsymbol{\Omega}}$ on $\mathcal{H}$, the norm $\|\cdot\|_\mathcal{H}$ permits defining a uniform metric, in which the distance between $M(k_1, -)$ and $M(k_2, -)$ is $\sup_{\omega \in \tilde{\boldsymbol{\Omega}}} \|M(k_1, \omega) - M(k_2, \omega)\|_\mathcal{H}$. Although in this paper we make no use of it, we define a metric deviation for POVMs $M$ and $M'$ that share the same knob domain $\boldsymbol{K}$ but can differ in both their Hilbert spaces and their detector domains as:

$$\mathrm{MetDev}(M, M') \stackrel{\mathrm{def}}{=}$$
$$\sup_{k_1, k_2 \in \boldsymbol{K}} \left| \sup_{\omega \in \tilde{\boldsymbol{\Omega}}} \| M(k_1, \omega) - M(k_2, \omega) \|_{\mathcal{H}} - \sup_{\omega' \in \tilde{\boldsymbol{\Omega}}'} \| M'(k_1, \omega') - M'(k_2, \omega') \|_{\mathcal{H}'} \right|. \tag{11}$$

## 4 Topologies and metrics induced on knobs by detections

The diversity of explanations available for any given statement of results prompts the question: what can we learn about knobs just from detection results $\mu$, without invoking any of the explanations in $\mathrm{Tr}^{-1}(\mu)$? So far knob domains have lacked topology. As outlined in Appendix B.1, for any set $\boldsymbol{K}$, a function from $\boldsymbol{K}$ to a metric space induces a topology on $\boldsymbol{K}$. Now probability measures on $\tilde{\boldsymbol{\Omega}}$ come with the uniform bounded metric $D_{\boldsymbol{\Omega}}$ defined in Eq. (3). View any parametrized probability measure $\mu$ as a function $\mu \colon \boldsymbol{K} \to \mathrm{PrMeas}(\tilde{\boldsymbol{\Omega}})$, where $\boldsymbol{K}$ is taken as a set, without any assumption of a topology, and $\mathrm{PrMeas}(\tilde{\boldsymbol{\Omega}})$ has the metric topology induced by $D_{\boldsymbol{\Omega}}$. For $V \subset \mathrm{PrMeas}(\tilde{\boldsymbol{\Omega}})$, define $\mu^{-1}(V) \stackrel{\mathrm{def}}{=} \{k \in \boldsymbol{K} | \mu(k, -) \in V\}$. Per Appendix B.1, $\mu$ induces a topology on $\boldsymbol{K}$ specified by

$$\tau_\mu = \{U \subset \boldsymbol{K} | (\exists V \text{ open in } \mathrm{PrMeas}(\tilde{\boldsymbol{\Omega}})) \quad U = \mu^{-1}(V)\}. \tag{12}$$

If $\mu$ is an injection into $\mathrm{PrMeas}(\tilde{\boldsymbol{\Omega}})$, then the (bounded) uniform metric on $\mathrm{PrMeas}(\tilde{\boldsymbol{\Omega}})$ induces a bounded metric on $\boldsymbol{K}$. If it is not injective, then $\mu$ induces a bounded metric on the quotient set of equivalence classes $\boldsymbol{K}/E_\mu$ where $E_\mu$ is the equivalence relation defined by

$$k_1 E_\mu k_2 \Leftrightarrow \mu(k_1, -) = \mu(k_2, -). \tag{13}$$

Examples of equivalence classes of knobs relevant to entangled states are discussed in [5], where they are level sets relevant to entangled states that violate Bell inequalities. When $\mu$ is not injective, in many cases the coarse topology $\tau_\mu$ on $\boldsymbol{K}$ induced by $\mu$ can be replaced by a finer topology by recognizing a finer level of description that augments the detector domain by adding another detector, as discussed below in connection with equivalence classes that characterize invariance.

**Remark**: Consider two parametrized probability measures $\mu$ and $\mu'$ having the same knob domain $\boldsymbol{K}$ but possibly distinct detector domains $\tilde{\boldsymbol{\Omega}}$ and $\tilde{\boldsymbol{\Omega}}'$, respectively. If their metric deviation is zero, then Appendix B.1 shows they induce the same topological and metric structures on $\boldsymbol{K}$.

$$\mathrm{MetDev}(\mu, \mu') = 0 \Rightarrow \tau_\mu = \tau_{\mu'}. \tag{14}$$

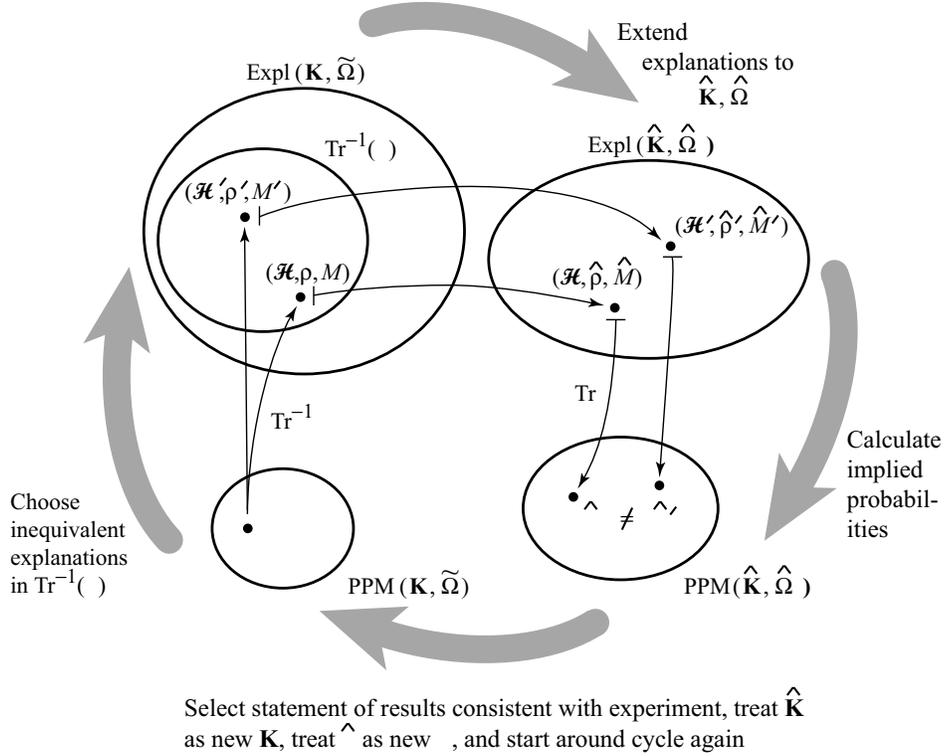

Figure 2: Expanding cycle of results and explanations.

## 5 Inequivalent explanations force extensions of domains

Although ambiguities preclude logic from forcing a single explanation, the existence of ambiguity logically forces a dynamic that continually extends statements of results and explanations. Picture a "penguin" toy walking down a slope with a rolling gait, leaning left and swinging its right leg, then leaning right and swinging its left leg, on and on in a cycle. From the proofs in [4] and the lattice structure of knob domains and detector domains follows the openness of a cycle of stating experimental results and explaining these results, as illustrated in Fig. 2. This cycle operates in a context not limited to theory but including the experimental endeavors that theory describes. While here we cannot reach beyond quantum formalism to touch them, we have experiments in mind as a background against which a statement of results implied by an explanation can be judged and, if incompatible, rejected.

Here is how the expanding cycle works. Given any $\boldsymbol{K}$ of more than one element and a generic $\mu \colon \boldsymbol{K} \to \mathrm{PrMeas}(\tilde{\boldsymbol{\Omega}})$ there are explanations $(\mathcal{H}, \rho, M), (\mathcal{H}', \rho', M') \in \mathrm{Tr}^{-1}(\mu)$ with the property that $\mathrm{MetDev}(\rho, \rho') \neq 0$ [4]; without loss of

generality suppose that

$$\tfrac{1}{2}\mathrm{Tr}_{\mathcal{H}}|\rho(k_1) - \rho(k_2)| > \tfrac{1}{2}\mathrm{Tr}_{\mathcal{H}'}|\rho'(k_1) - \rho'(k_2)|. \tag{15}$$

Suppose the explanation is expanded to cover a larger knob domain in such a way that the density operator and the POVM can be independently selected. Leaving the choice of density operators unchanged, consider the effect of the availability of certain additional POVMs as expressed by expanding the domain of knobs from $\boldsymbol{K}$ to $\hat{\boldsymbol{K}} = \boldsymbol{K} \vee \boldsymbol{L}$ where $\boldsymbol{L}$ comprises a copy of $\boldsymbol{K}$ and, in addition, an extra knob domain $\boldsymbol{B}$. The explanations expand to $(\mathcal{H}, \hat{\rho}, \hat{M}), (\mathcal{H}', \hat{\rho}', \hat{M}')$ in which the POVMs $\hat{M}$ and $\hat{M}'$ are functions on $\boldsymbol{L} \cong \boldsymbol{K} \vee \boldsymbol{B}$. We design one setting $b_0 \in \boldsymbol{B}$ to work so that $(\forall k \in \boldsymbol{K})$ the setting $(k, k, b_0) \in \hat{\boldsymbol{K}}$ has the same effect as does $k \in \boldsymbol{K}$; that is, the expanded explanations envelop the given explanations by the mappings

$$\hat{M}(k, b_0, -) = M(k, -), \tag{16}$$
$$\hat{M}'(k, b_0, -) = M'(k, -). \tag{17}$$

Correspondingly, the statements of results implied by the explanations are also enveloped [5] by the larger domain under the condition that $b = b_0$, in that we have $(\forall k \in \boldsymbol{K})$

$$\begin{aligned}\hat{\mu}(k, k, b_0, \omega) &= \mathrm{Tr}_{\mathcal{H}}[\rho(k)\hat{M}(k, b_0, \omega)] \\ &= \mathrm{Tr}_{\mathcal{H}}[\rho(k)M(k, \omega)] = \mu(k, \omega), \end{aligned} \tag{18}$$
$$\begin{aligned}\hat{\mu}'(k, k, b_0, \omega') &= \mathrm{Tr}_{\mathcal{H}'}[\rho'(k)\hat{M}'(k, b_0, \omega')] \\ &= \mathrm{Tr}_{\mathcal{H}'}[\rho(k)M'(k, \omega')] = \mu'(k, \omega'). \end{aligned} \tag{19}$$

For another setting $b_1$ of $\boldsymbol{B}$, however, the expanded explanations can be chosen to conflict in the results they imply. In particular, we are free to choose an explanation in which, independent of $k$, $\hat{M}(k, b_1)$ is the optimal POVM for deciding between $\rho(k_1)$ and $\rho(k_2)$, while, also independent of $k$, we take $\hat{M}'(k, b_1)$ to be the optimal POVM for deciding between $\rho'(k_1)$ and $\rho'(k_2)$. The trace functor generates the corresponding probability measures on the extended knob domain:

$$\hat{\mu}(k_j, k, b_1, \omega) = \mathrm{Tr}_{\mathcal{H}}[\rho(k_j)\hat{M}(k, b_1, \omega)], \tag{20}$$
$$\hat{\mu}'(k_j, k, b_1, \omega') = \mathrm{Tr}_{\mathcal{H}'}[\rho'(k_j)\hat{M}'(k, b_1, \omega')], \tag{21}$$

where $j \in \{1, 2\}$. Subtracting the case $j = 2$ from the case $j = 1$ yields

$$\begin{aligned}|\hat{\mu}(k_1, k, b_1, \omega) - \hat{\mu}(k_2, k, b_1, \omega)| &= \left|\mathrm{Tr}_{\mathcal{H}}[\hat{M}(k, b_1, \omega)\{\rho(k_1) - \rho(k_2)\}]\right| \\ &= \tfrac{1}{2}\mathrm{Tr}_{\mathcal{H}}|\rho(k_1) - \rho(k_2)|, \end{aligned} \tag{22}$$

where the last equality follows from a property of optimal decision operators [10]. For the "primed" extended explanation the same logic implies

$$\begin{aligned}|\hat{\mu}'(k_1, k, b_1, \omega') - \hat{\mu}'(k_2, k, b_1, \omega')| &= \tfrac{1}{2}\mathrm{Tr}_{\mathcal{H}'}|\rho'(k_1) - \rho'(k_2)| \\ &\neq \tfrac{1}{2}\mathrm{Tr}_{\mathcal{H}}|\rho(k_1) - \rho(k_2)|, \end{aligned} \tag{23}$$

whence the extensions of the metrically inequivalent explanations $(\mathcal{H}, \rho, M), (\mathcal{H}', \rho', M') \in \text{Tr}^{-1}(\mu)$, for which $\text{MetDev}(\rho, \rho') \neq 0$, though they agree in their implied results on $\boldsymbol{K}$, have extensions that imply metrically inequivalent statements of results $\hat{\mu}$ and $\hat{\mu}'$ on $\boldsymbol{K} \vee \boldsymbol{L}$ and so conflict in their predictions of knob physics.

The upshot is an open cycle. Expanding the knob domain allows extensions of metrically inequivalent explanations to imply conflicting extended statements of results $\hat{\mu}$ and $\hat{\mu}'$. On rejecting one of these statements, say on the basis of experiment, and assuming the other statement of results, one treats $\hat{\boldsymbol{K}}$ as a new "given knob domain $\boldsymbol{K}$," and the cycle starts over, but cycling round investigations of ever-expanding knob domains.

A lesson on the negative side is this. If quantum descriptions are inherently multiple, look for trouble in endeavors that assume the conceptual possibility of a single explanation. For example, many investigators of quantum key distribution [11] have failed to disentangle quantum decision theory from a self-contradictory (usually unspoken) assumption that a single explanation makes sense. Although quantum key distribution avoids many of the vulnerabilities of classical key transmission, some new potential vulnerabilities arise, at root because of the multiplicity of explanations with their conflicting extensions to a larger knob domain. The conflicting extensions mean that no single explanation can logically substitute for an experimental investigation of that larger domain. For a sketch of the details see Appendix D. On the positive side, acceptance of descriptions as inherently multiple resolves a conceptual muddle, to which we now turn.

## 6 Making sense of invariance

To demonstrate an invariance we might place a round drinking glass on a table and rotate it to show that "nothing changes under rotation." But to see this invariance, whether one is aware of it or not, one must manage incompatible frames of reference [12]. Looked at one way "nothing happens when we rotate the glass"; but to see that "nothing happens when we rotate the glass" one must see in the other frame, so to speak, that in fact "the glass rotates," as evidenced perhaps by a visible speck of dust on the glass or other irregularity that, strictly speaking, violates its symmetry and thereby makes visible its rotation.

Formally, an invariance shows up in a statement of results $\mu \colon \boldsymbol{K} \to \text{PrMeas}(\tilde{\Omega})$ as an equivalence relation $E_\mu$ on $\boldsymbol{K}$, with equivalence classes

$$E_\mu(k) \stackrel{\text{def}}{=} \{k' | \mu(k', -) = \mu(k, -)\}.$$

Changing $k$ within an equivalence class—a level set—leaves all the probabilities of detections invariant. A colleague at an earlier talk asked "then why not 'mod out' the equivalence classes?" Indeed, if certain changes of knob settings make no difference, what experimental evidence do we have to speak of 'changing a knob setting' at all?

Yet physicists need to speak of changes that "don't do anything." For example, special relativity deals with how nothing changes when a train is put in uniform motion relative to a station. But, just as in the water glass, the "nothing changes" must be seen also from a conflicting viewpoint in which "the train moves." If one tries to condense the concept of invariance into a single description, one meets confusion; while if we recognize that multiplicity of description as part and parcel of the concept of description, two levels of description suffice to make invariance comprehensible.

Here is an example involving mapping a detector domain $\tilde{\mathbf{\Omega}}$ into a larger detector domain $\tilde{\mathbf{\Omega}}' = \tilde{\mathbf{\Omega}} \vee \tilde{\mathbf{\Omega}}''$, where $\tilde{\mathbf{\Omega}} \wedge \tilde{\mathbf{\Omega}}'' = \emptyset$; in effect $\tilde{\mathbf{\Omega}}'$ augments $\tilde{\mathbf{\Omega}}$ with an additional detector. The mapping is the injection $g : \tilde{\mathbf{\Omega}} \rightarrowtail \tilde{\mathbf{\Omega}}'$ that assigns to each $(\omega)$ in the smaller detector domain $\tilde{\mathbf{\Omega}}$ the element $(\omega, \Omega'') \in \tilde{\mathbf{\Omega}}'$. (In effect, $g(\omega)$ ignores the detectors of $\tilde{\mathbf{\Omega}}'$ other than those expressed by $\tilde{\mathbf{\Omega}}$.) The injection $g$ induces a "contravariant" map $F_g : \text{PPM}(\boldsymbol{K}, \tilde{\mathbf{\Omega}}') \to \text{PPM}(\boldsymbol{K}, \tilde{\mathbf{\Omega}})$ that corresponds to marginal probability; that is $F_g$ is defined by

$$F_g(\mu') = \mu \text{ s.t. } (\forall k \in \boldsymbol{K}, \omega \in \tilde{\mathbf{\Omega}}) \quad \mu(k, \omega) = \mu'(k, g(\omega)) = \mu'(k, (\omega, \Omega'')). \quad (24)$$

The extra detail needed to "see the glass move" shows up in the finer-level statement of results $\mu'$, with its dependence on an extra detector that, like the speck of dust, expresses what happened that left the marginal probabilities $\mu$ invariant. Taking the marginal parametrized probability obtained by ignoring the extra detector, we get $\mu'[k, (\omega, \Omega'')] = \mu(k, \omega)$, so the invariance in $\mu$ is retrieved, now seen as a *marginal* parametrized probability measure, derived from a more complex parametrized probability measure $\mu'$.

## 7   Discussion

In broad terms, we offer here a recognition of guesswork as a third pillar of science along side of calculation and measurement. Among the giants on whose shoulders we stand are Ernst Mach and those of his intellectual descendants, Heisenberg among them, who worried that theory seemed so remote from life in a laboratory equipped with instruments of measurement. We owe a debt to both the push toward operationalism and the counter-push that recognizes the need for theoretical constructs having no direct counterparts on the lab bench. Tracing out the history of these ideas in relation to an acceptance of ambiguity and the consequent role of guesswork is an appealing project for a future collaboration with the historically literate.

We introduce the unfamiliar term *explanation* for the vectors and operators of a quantum description to contrast them with the probabilities which we say are explained. This contrast is plain enough to see in the texts of Dirac and von Neumann, etc. but we aimed to highlight it. Striking to us is the way quantum theory provides both structure and ambiguity, and the structure of probabilities becomes an essential ingredient in ambiguity of outcomes, as here conceived.

We were asked about the relationship of our work to elements of quantum logic and quantum probability theory. Quantum logic expresses measurement outcomes as subspaces of a Hilbert space, while we express measurement outcomes as a field of sets, in the sense of Kolmogorov [13], without any reference to a Hilbert space. By detaching our concept of an event from a Hilbert space, we acquire the power to speak of differing explanations involving differing Hilbert spaces that "explain" the same probabilities of outcomes. Within any single explanation, the outcomes explained can be made to correspond to subspaces of a Hilbert space (perhaps with the use of Neumark's theorem [14] to convert an arbitrary POVM to a projective POVM).

In connection with quantum logic, we were asked about the relation of our work to that of Birkhoff and von Neumann [15], who showed a role for two distinct lattices, one to do with their propositional calculus, the other to do with subspaces of a Hilbert space. We are on the side of those who notice that in their demonstrations of lattice properties, Birkhoff and von Neumann use garden-variety propositional logic, and not their quantum logic, which to us is an interesting display of lattice structures. Naming the lattice structure of Hilbert spaces "quantum logic" seems to us a misnomer that confuses the unwary. As amateurs looking from "outside," we enjoy the disparate views among mathematicians and logicians about suitable rules for the game of logic; we have not yet seen anything that can reasonably be termed "quantum" about logic.

We were asked also about the relation of our work to Gleason's theorem. The short answer is that we notice that in physics the Hilbert space is never an experimental fact, but requires an act of guesswork. Once one guesses a Hilbert space, then certainly Gleason's theorem comes into play, and our work is consistent with it, because we pay attention to knob domains that are roughly speaking smaller than the space of operators on a Hilbert space. By way of justification for the focus on knob domains that are "small" relative to a space of operators, note that choosing an explanation involving a Hilbert space to which Gleason's theorem pertains, and taking the knob domain to be the whole space of density operators and POVMs on that Hilbert space, there is always an alternative explanation involving a larger Hilbert space with its larger space of operators, relative to which that knob domain becomes "small."

# Acknowledgments

We are grateful for the invitation to present a preliminary version of this paper at the Conference on Representation Theory, Quantum Field Theory, Category Theory, Mathematical Physics and Quantum Information Theory, funded by the National Science Foundation, September 20–23, 2007, at The University of Texas at Tyler, organized by Kazem Mahdavi, Louis Kauffman, Samuel Lomonaco, and Deborah Koslover. We also thank Sam Lomonaco for helpful suggestions. We thank an anonymous referee for posing the questions to which we respond in the Discussion section.

# A  Unordered products of knobs and detectors

For any set $X$ of sets, the unordered product $\boldsymbol{\pi} X$ is a set of pairs, with each pair of the form $(a, A)$, where $a \in A$, with exactly one pair for each set in $X$. E.g., if $X = \{A, B, C\}$, then $\boldsymbol{\pi} X = \{\,\{(a, A), (b, B), (c, C)\} | a \in A, b \in B, c \in C\}$. We call the sets $A, B, C \in X$ *factors* of the unordered product $\boldsymbol{\pi} X$. Unordered products of sets constitute a lattice with a partial order, join, meet and difference defined by

$$\boldsymbol{\pi} X \leq \boldsymbol{\pi} Y \stackrel{\text{def}}{=} X \subset Y, \tag{25}$$

$$\boldsymbol{\pi} X \vee \boldsymbol{\pi} Y \stackrel{\text{def}}{=} \boldsymbol{\pi}(X \cup Y), \tag{26}$$

$$\boldsymbol{\pi} X \wedge \boldsymbol{\pi} Y \stackrel{\text{def}}{=} \boldsymbol{\pi}(X \cap Y), \tag{27}$$

$$\boldsymbol{\pi} X \stackrel{\boldsymbol{\pi}}{-} \boldsymbol{\pi} Y \stackrel{\text{def}}{=} \boldsymbol{\pi}(X - Y), \tag{28}$$

where $X - Y$ is the set difference: $X - Y \stackrel{\text{def}}{=} \{a | a \in X \text{ and } a \notin Y\}$.

**Definition**: Understanding a *knob* to be a set of knob settings, a *knob domain* is an unordered product of knobs. We call an element of a knob domain a setting (of that domain); it conveys the settings of all the knobs of the domain.

**Remarks**:

1. A knob domain resembles a cartesian product of knobs, except that: (a) it excludes the possibility of the same knob appearing twice; and (b) t avoids the ordering presumed by a cartesian product. Both of these exceptions to the cartesian product are required for a *join* of two knob domains, defined below, to make sense.

2. Underlying any knob domain $\boldsymbol{K}$ is its set of knobs which we denote by $\boldsymbol{\pi}^{-1} \boldsymbol{K}$. This $\boldsymbol{\pi}^{-1}$ is a forgetful functor from unordered products to their underlying sets. It follows that

    $$(\text{for } X \text{ any set of sets}) \quad \boldsymbol{\pi}^{-1}(\boldsymbol{\pi} X) = X, \tag{29}$$

    and

    $$(\text{for } \boldsymbol{K} \text{ any knob domain}) \quad \boldsymbol{\pi}(\boldsymbol{\pi}^{-1} \boldsymbol{K}) = \boldsymbol{K}. \tag{30}$$

Their definition as unordered products implies that a set of knob domains has a lattice structure, handy for expressing the relation between two experiments that share some but not all of the same knobs. Given two knob domains $\boldsymbol{K}$ and $\boldsymbol{K}'$, their meet $\boldsymbol{K} \wedge \boldsymbol{K}'$ amounts to the knobs they share in common, while their join is related to the join of their underlying unordered products.

By way of the forgetful functor $\boldsymbol{\pi}^{-1}$ that takes an unordered product to the set of its underlying sets, this lattice of knob domains is defined, for any two knob domains $K$ and $K'$, by the following:

$$K \leq K' \stackrel{\text{def}}{=} \pi^{-1}(K) \subset \pi^{-1}(K'), \tag{31}$$

$$K \vee K' \stackrel{\text{def}}{=} \pi[\pi^{-1}(K) \cup \pi^{-1}(K')], \tag{32}$$

$$K \wedge K' \stackrel{\text{def}}{=} \pi[\pi^{-1}(K) \cap \pi^{-1}(K')], \tag{33}$$

$$K' \stackrel{\pi}{-} K \stackrel{\text{def}}{=} \pi[\pi^{-1}(K') - \pi^{-1}(K)]. \tag{34}$$

If two knob domains share no common knob in their underlying sets, we have $K \wedge K' = \emptyset$. In case $K \wedge K' = \emptyset$ (and only in this case), an isomorphism takes any $(x, y)$ with $x \in K$ and $y \in K'$ to an element $z \in K \vee K'$. We express this isomorphism as

$$z = x \vee y. \tag{35}$$

Two distinct ways of getting less than a knob domain play a role. Given a knob domain $K'$, we can be interested in a domain $K$ that has some but not all of the same knobs, a relation written as $K < K'$. Or, we can be interested in a subset of $L \subsetneq K'$. Each element of $L$ specifies a setting of each knob of $K'$ but some of the elements of $K'$ are absent from $L$.

Correspondingly, two levels of set differences enter the story. If $L \subset K$, then $K - L \stackrel{\text{def}}{=} \{x | x \in K \text{ and } x \notin L\}$. This is the ordinary set difference. We also want another kind of difference that applies to two knob domains $K$ and $K'$. If the two knob domains do not share all the same knobs, they have no elements in common (so that $K' - K = K'$); we use $\stackrel{\pi}{-}$ in the expression $K' \stackrel{\pi}{-} K$ to indicate the knob domain is an unordered product of those factors of $K'$ that are not factors of $K$.

## A.1 Detector domains

We suppose that each detector separately is expressed mathematically by an outcome space. To each outcome space $\Omega$ there is associated a set $\tilde{\Omega}$ of the measurable subsets of $\Omega$. Now extend this construction by letting $\Omega$ be the unordered product of a set of outcome spaces $\{\Omega_A, \Omega_B, \ldots\}$. Let $\tilde{\Omega}$ be the set of measurable subsets of this product, constructed in analogy with the construction for cartesian products of measure spaces, and call such an entity a *detector domain*. A detector domain $\tilde{\Omega}$ built up from more than one outcome set contains, in addition to unordered measurable rectangles, unions of disjoint measurable rectangles; indeed it is defined to be a $\sigma$-algebra [16].

In parallel with the story for knobs, for detectors we have a lattice of unordered products of outcome spaces. This lattice of unordered products of outcome spaces induces a lattice of detector domains defined by

$$\tilde{\Omega} \leq \tilde{\Omega}' \stackrel{\text{def}}{=} \widetilde{\Omega \leq \Omega'}, \tag{36}$$

$$\tilde{\Omega} \vee \tilde{\Omega}' \stackrel{\text{def}}{=} \widetilde{\Omega \vee \Omega'}, \tag{37}$$

$$\tilde{\Omega} \wedge \tilde{\Omega}' \stackrel{\text{def}}{=} \widetilde{\Omega \wedge \Omega'}, \tag{38}$$

$$\tilde{\Omega}' \stackrel{\pi}{-} \tilde{\Omega} \stackrel{\text{def}}{=} \widetilde{\Omega' \stackrel{\pi}{-} \Omega}. \tag{39}$$

# B Metric deviation of two functions having the same domain but possibly distinct codomains

Consider any two spaces $Y$ and $Y'$ equipped with bounded metrics $d$ and $d'$, respectively; let $X$ be any set of more than one element. We define the *metric deviation* between any two functions $f : X \to Y$ and $f' : X \to Y'$ by

$$\text{MetDev}(f, f') \stackrel{\text{def}}{=} \sup_{x_1, x_2 \in X} |d(f(x_1), f(x_2)) - d'(f'(x_1), f'(x_2))|. \tag{40}$$

If the deviation is zero, we speak of $f$ and $f'$ as *metrically equivalent.*

In the case $Y = Y'$, the functions $f$ and $f'$ can also be compared by the uniform metric. In this case the metric deviation is a coarser comparison than the uniform metric, and it is this coarser comparison that is most relevant to what detections can tell about the structure of knob domains.

## B.1 Topology induced on a domain by a function to a space

Let $Y$ be any topological space with topology $\tau_Y$, let $X$ be any set, and let $f$ be any function from the domain $X$ to $Y$. Then $f$ partitions $X$ into equivalence classes by the relation

$$x E_f x' \Leftrightarrow f(x) = f(x'). \tag{41}$$

Let $X/E_f$ be the quotient set (set of equivalence classes) of $X$ by $E_f$, and define the projection

$$p_f : X \to X/E_f. \tag{42}$$

1. Define a topology $\tau_f$ on the domain $X$ by

$$\tau_f = \{U | (\exists V \in \tau_Y) \quad U = f^{-1}(V)\}. \tag{43}$$

2. Consider $X/E_f$ with the topology $p_f \tau_f \stackrel{\text{def}}{=} \{W | (\exists U \in \tau_f) \ W = p_f U\}$. $X/E_f$ with this topology is homeomorphic to $\text{Im } f$ with its subspace topology inherited from $\tau_Y$.

3. Any metric $d$ on $Y$ induces a metric $d_f$ on $X/E_f$:

$$d_f(p_f x_1, p_f x_2) \stackrel{\text{def}}{=} d[f(x_1), f(x_2)]. \tag{44}$$

4. If $Y$ has a metric $d$ and $f$ is injective, then $d_f$ is a metric on $X$.

**Lemma**: For $f : X \to Y$ and $f' : X \to Y$, if $\text{MetDev}(f, f') = 0$ then

$$\tau_f = \tau_{f'} \text{ and } d_f = d_{f'}. \tag{45}$$

*Proof*: $\text{MetDev}(f, f') = 0 \Rightarrow E_f = E_{f'}$, whence follows $\tau_f = \tau_{f'}$. □

# C  Diverse explanations of given results

Although statements of results leave open choices of explanations, they do indeed impose some constraints. Here we show that except in limiting special cases these constraints leave metrically inequivalent density-operator functions available for explanations within $\mathrm{Tr}^{-1}(\mu)$.

In [4] we rather arbitrarily imposed an additional constraint on explanations (which in that paper we called *models*) by separating control over the density operator from control over the POVM, so that a knob domain has the form $\boldsymbol{K} = \boldsymbol{A} \times \boldsymbol{B}$, with

$$(\forall k \in \boldsymbol{K})(\exists a \in \boldsymbol{A}, b \in \boldsymbol{B}) \quad k = (a,b). \tag{46}$$

From Propositions 2 and 4 of [4] follows the

**Proposition**: Even under this additional constraint on explanations, there are metrically inequivalent density operators in $\mathrm{Tr}^{-1}(\mu)$ *unless*

$$(\forall a_1, a_2 \in \boldsymbol{A})(\exists b \in \boldsymbol{B}) \quad D_{\boldsymbol{\Omega}}[\mu(a_1,b), \mu(a_2,b)] = 1. \tag{47}$$

## C.1  Constraint on explanations imposed by given results

A simpler if weaker demonstration of constraints imposed by results than that given in [4] is the following. A given statement of results $\mu \in \mathrm{PPM}(\boldsymbol{K}, \tilde{\boldsymbol{\Omega}})$ imposes some constraints on explanations, as follows. Using the definition of $D_{\boldsymbol{\Omega}}$ given in Eq. (4),

$$\begin{aligned}
\big(\forall (\mathcal{H}, \rho, M) &\in \mathrm{Tr}^{-1}(\mu)\big) \quad D_{\boldsymbol{\Omega}}[\mu(k_1,-), \mu(k_2,-)] \\
&= \sup_{\omega \in \tilde{\boldsymbol{\Omega}}} \mathrm{Tr}_{\mathcal{H}}[\rho(k_1) M(k_1, \omega) - \rho(k_2) M(k_2, \omega)] \\
&= \sup_{\omega \in \tilde{\boldsymbol{\Omega}}} \big(\mathrm{Tr}_{\mathcal{H}}\{[\rho(k_1) - \rho(k_2)] M(k_1, \omega)\} + \mathrm{Tr}_{\mathcal{H}}\{\rho(k_2)[M(k_1, \omega) - M(k_2, \omega)]\}\big) \\
&\leq \sup_{M' \in \mathrm{DetectOp}(\mathcal{H})} \mathrm{Tr}\{M'[\rho(k_1) - \rho(k_2)]\} \\
&\quad + \sup_{\rho' \in \mathrm{DensOp}(\mathcal{H})} \mathrm{Tr}_{\mathcal{H}}\{\rho'[M(k_1, \omega) - M(k_2, \omega)]\} \\
&= \tfrac{1}{2} \mathrm{Tr}_{\mathcal{H}} |\rho(k_1) - \rho(k_2)| + \sup_{\omega \in \tilde{\boldsymbol{\Omega}}} \|M(k_1, \omega) - M(k_2, \omega)\|_{\mathcal{H}}. \tag{48}
\end{aligned}$$

The last equality makes use of the relation shown in [10] for trace distance and also hermitian property of detection operators.

# D  Ambiguity of explanations in quantum cryptography

In quantum cryptography, specifically quantum key distribution, untenable claims of absolute security against undetected eavesdropping have arisen from the tacit supposition of a single explanation of experimental results. Under that supposition, security claims invoke a theorem of quantum decision theory that tells how the minimum probability of error for deciding between two states $\rho(1)$ and $\rho(2)$ rises as their trace distance decreases. For example, without regard to the multiplicity of explanations available for any given probabilities, the popular design BB84 [17] invokes a single explanation in which pairs of states $\rho(1)$ and $\rho(2)$ exhibit a trace distance less than or equal to $2^{-1/2}$, implying a minimum probability of error to decide between them:

$$P_E \geq \tfrac{1}{2}(1 - \tfrac{1}{2}|\rho(1) - \rho(2)|) = \tfrac{1}{2}(1 - \sqrt{\tfrac{1}{2}}) \approx 0.146. \tag{49}$$

But how is one to rely on an implemented key-distribution system built from lasers and optical fibers and so forth to act in accordance with this explanation? If a system of lasers and optical fibers and so forth "possessed" a single explanation in terms of quantum states, one could hope to test experimentally the trace distance between the pair of states. But no such luck. The trouble is that trace distance is a property not of probabilities *per se*, which are testable, but of some one among the many *explanations* of those probabilities. While the testable probabilities constrain the possible explanations, and hence constrain trace distances, this constraint on trace distance is "the wrong way around"—a lower bound instead of a sub-unity upper bound on which security claims depend.

Given any parametrized probability measure, proposition 2 in [4] assures the existence of an explanation in terms of a parametrized density operator $\rho'$ metrically inequivalent to $\rho$, such that, in conflict with Eq. (49), the trace distance becomes $\tfrac{1}{2}|\rho'(1) - \rho'(2)| = 1$, making the quantum states in this explanation distinguishable without error, so that the keys that they carry are totally insecure.

The central issue in key distribution is this: how will the lasers and fibers and detectors that convey the key respond to attacks, in which an as yet unknown eavesdropper brings extra devices with their own knobs and detectors into contact with the key-distributing system? Attacks entail knob and/or detector domains extended beyond those tested, with the possibility that extended explanations metrically inequivalent to that used in the design, but consistent with available probabilities, both imply a lack of security theoretically and accord with actual eavesdropping.

Physically, one way for insecurity to arise is by an information leak through frequency side-band undescribed in the explanation on which system designers relied. A more likely security hole appears when lasers that are intended to radiate at the same light frequency actually radiate at slightly different frequencies, as described in [5, 18, 19].